# Parallel-plate chambers as radiation-hard detectors for time-based beam diagnostics in carbon-ion radiotherapy

Na Hye Kwon[1], Sung Woon Choi[1], Soo Rim Han[1], Yongdo Yun[1], Min Cheol Han[1], Chae-Seon Hong[1], Ho Jin Kim[1], Ho Lee[1], Changhwan Kim[1], Do Won Kim[1], Woong Sub Koom[1], Jin Sung Kim[1,2], N. Carolino[3], L. Lopes[3], Dong Wook Kim[1,5]\*, Paulo J. R. Fonte[1,3,4]\*

[1] Department of Radiation Oncology, Yonsei Cancer Center, Heavy Ion Therapy Research Institute, Yonsei University College of Medicine, Seoul, Republic of Korea

[2] Oncosoft Inc., Seoul, South Korea

[3] LIP – Laboratory of Instrumentation and Experimental Particle Physics, Coimbra, Portugal

[4] Coimbra Institute of Engineering, Polytechnic University of Coimbra, Portugal

[5] ARA Inc., Seoul, South Korea

\* These authors contributed equally to this work.
\* joocheck@gmail.com (DW Kim); fonte@lip.pt (Paulo Fonte)

## Abstract

Accurate range verification of carbon ion beams is critical for the precision and safety of charged particle radiotherapy. In this study, we evaluated the feasibility of using a parallel-plate ionization chamber for real-time, time-based diagnostic monitoring of carbon ion beams. The chamber featured a 0.4 mm gas gap defined by metallic electrodes and was filled with carbon dioxide ($CO_2$), a non-polymerizing gas suitable for high-rate applications. Timing precision was assessed via self-correlation analysis, yielding a precision approaching one picosecond for one-second acquisitions under clinically relevant beam conditions. This level of timing accuracy translates to a water-equivalent range uncertainty of approximately 1 mm, which meets the recommended clinical tolerance for carbon ion therapy. Furthermore, the kinetic energy of the beam at the synchrotron extraction point was determined from the measured orbital period, with results consistently within 1 MeV/nucleon of the nominal energy. These findings demonstrate the potential of parallel-plate chambers for precise, real-time energy and range verification in clinical carbon ion beam quality assurance.

# 1. Introduction

Particle therapy, particularly with protons and carbon ions, enables highly conformal tumor irradiation by exploiting the steep dose falloffs after the Bragg peak. This precision in dose delivery demands stringent quality assurance (QA) protocols to ensure both the accuracy and reproducibility of treatments [1–3]. To address these requirements, the American Association of Physicists in Medicine (AAPM) has published comprehensive guidelines through reports such as Task Group (TG) 142 and 224, which are widely implemented in clinical particle therapy practice [4,5].

Modern radiotherapy QA encompasses daily, monthly, and annual procedures, as well as patient-specific QA performed prior to each treatment. These protocols are designed prevent delivery errors and to ensure that the prescribed dose is accurately administered to the target volume [6]. Among monthly QA tasks, range verification based on beam energy is particularly critical to achieve precise tumor coverage while minimizing dose exposure to surrounding healthy tissue [7]. A commonly employed technique utilizes commercial multilayer ionization chamber arrays, such as the ZEBRA and GIRAFFE detector (IBA Dosimetry, Germany), to measure the longitudinal depth-dose distribution of central-axis pencil beams [8,9].

Despite their clinical utility, conventional QA approaches are often time-consuming and labor-intensive, which can limit their practicality in fast-paced clinical environments. More importantly, these approaches are generally unable to detect transient range deviations that may arise during actual treatment delivery. To improve efficiency and enable verification of dose delivery, log-based QA systems have been introduced, offering three-dimensional recalculation capabilities for both physical and biological dose distributions [10]. While these systems have proven effective in streamlining patient-specific QA (PSQA) in CIRT, they are inherently incapable of directly detecting deviations in beam energy during treatment.

In particular, carbon ion beams may undergo unintended energy loss or scattering after extraction from the synchrotron, as they traverse nozzle components - including monitoring chambers, air gaps, or range shifters. Or, in some cases, carbon ion beams may interact with structures inside the nozzle, causing unintended energy loss. Crucially, such deviations remain undetected by standard machine logs or DICOM RT treatment records, which lack actual energy delivery data during irradiation [10,11]. This critical limitation underscores the need for an independent, real-time monitoring system capable of detecting beam energy variations at the point of delivery.

To address this clinical requirement, radiation-hard detectors enabling direct, high-speed beam energy monitoring during treatment are essential. Parallel-plate chambers (PPCs) represent a promising solution due to their high-rate capability, easily replaceable gas-based active medium, and radiation-resistant inert electrode materials [12-14]. However, technical challenges include limited dynamic range (with sparking risks under high rate conditions) and gas polymerization—which can deposit insulating materials on the electrodes and degrade performance.

In this study, we evaluated the feasibility of a specialized PPC detector for range verification in CIRT. We quantify its timing precision and demonstrate its feasibility for real-time clinical beam monitoring.

## 2. Materials and Methods

### 2.1. Development of the Parallel-Plate Chamber (PPC)

For feasibility assessment, we designed and constructed a parallel-plate chamber (PPC) with a circular active area of 2.5 cm in diameter and a 0.4 mm gas gap, as shown in Fig. 1 (a). Both the anode and cathode electrodes were fabricated from stainless-steel, chosen for mechanical robustness and ease of machining. In the active region the thickness of each electrode was 0.5 mm. The assembled PPC was centrally mounted within a gas-tight detector housing, as illustrated in Fig. 1 (b), and connected to a high-voltage (HV) power supply and signal readout electronics. The front and rear windows of the chamber enclosure were made of transparent Kapton film, which provided effective gas sealing while allowing the particle beam to traverse the chamber with minimal attenuation and scattering. A second window made of thin aluminium foil ensured a proper electrical shielding.

Carbon dioxide ($CO_2$) was selected as the operating gas due to its chemical stability and resistance to polymerization. This choice enhances the radiation hardness of the detector by preventing the formation of polymer deposits on the electrodes, which could otherwise induce Malter effect, leading to increased dark currents or even electrical breakdown (sparking). [15].

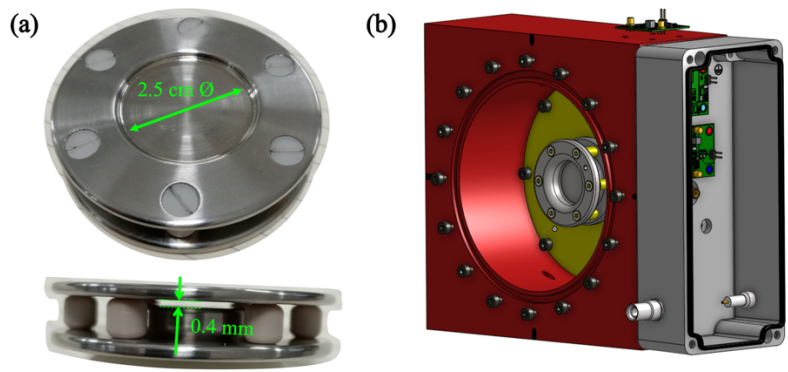

**Fig. 1.** (a) Constructed PPC detector with a 2.5 cm active diameter and 0.4 mm gas gap,
(b) CAD drawing of the detector housing with the PPC mounted at the center

#### 2.1.1. Configuration of the PPC and Electronics Readout

The HV distribution and signal readout circuits are illustrated in Fig. 2. The electronics chain comprises a spark protection circuit followed by two cascaded high-frequency amplifiers. A symmetric HV bias (±HV) is applied to the electrodes via a resistor–capacitor (RC) network, as depicted in Fig. 2 (a). Each electrode is connected through a pair of 1 MΩ current-limiting resistors (R2, R4) in series with an RC filter for rejection of the pickup noise that might be introduced from the HV lines.

The detector signal is extracted at point P1, located between the two 220 pF coupling capacitors (C3, C4), and routed to the external readout electronics. As shown in Fig. 2 (b), the signal passes from P2 to P3 through a dedicated spark protection circuit, which safeguards downstream electronics from voltage spikes and transients. The signal is then fed into the amplification chain, proceeding from P4 to P5, as illustrated in Fig. 2 (c). The amplification stage utilizes a monolithic microwave integrated circuit (MMIC) gain stage (Infineon BGA614) transformer-coupled to the next stage. The input impedance, defined by the MMIC, is 50 Ω. This configuration provides sufficient gain while preserving the fast-timing characteristics essential for high-resolution signal processing.

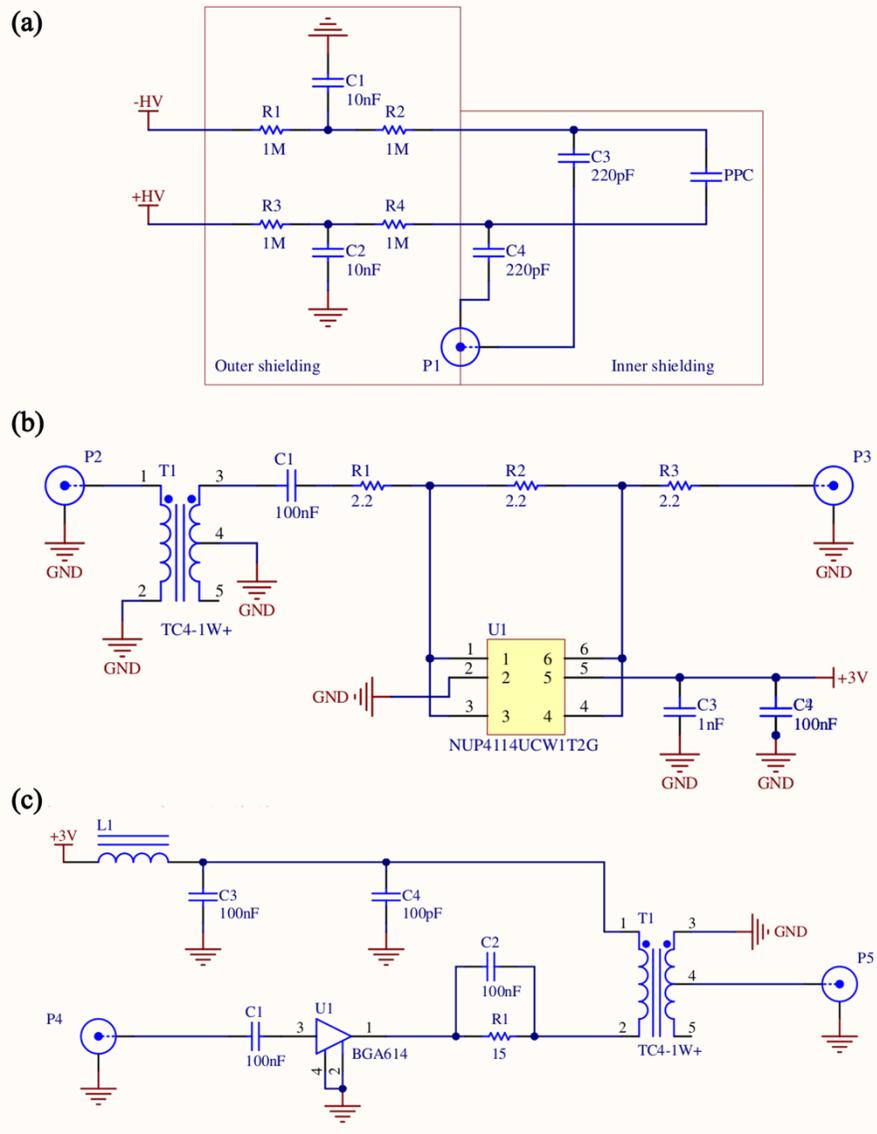

**Fig. 2.** (a) HV distribution and signal readout configuration of the PPC, (b) spark protection circuit, and (c) amplifier stages with transformer coupling (two of these are cascaded).

To assess the dynamic response of the readout electronics, the PPC was replaced by a calibrated current pulse source. As depicted in Fig. 3, the yellow trace represents the input pulse generated by a function generator (Model 4054B, BK Precision, USA), while the magenta trace shows the impulse response of the circuit after passing through the complete electronics chain including the spark protection circuit and two cascaded amplifier stages.

The injected pulse generated a current of approximately 100 µA peak value (equivalent to 500 mV across a 5000 Ω load) and the resulting output signal exhibited a peak amplitude of 370 mV, indicating a high-frequency transimpedance of 3.7 kΩ, corresponding to a voltage gain of 74. A noticeable undershoot was observed, primarily attributable to the 3 MHz low-frequency cutoff imposed by the coupling transformers and the 220 pF series capacitors. The high-frequency cutoff is 800 MHz, limited by the transformers bandwidth.

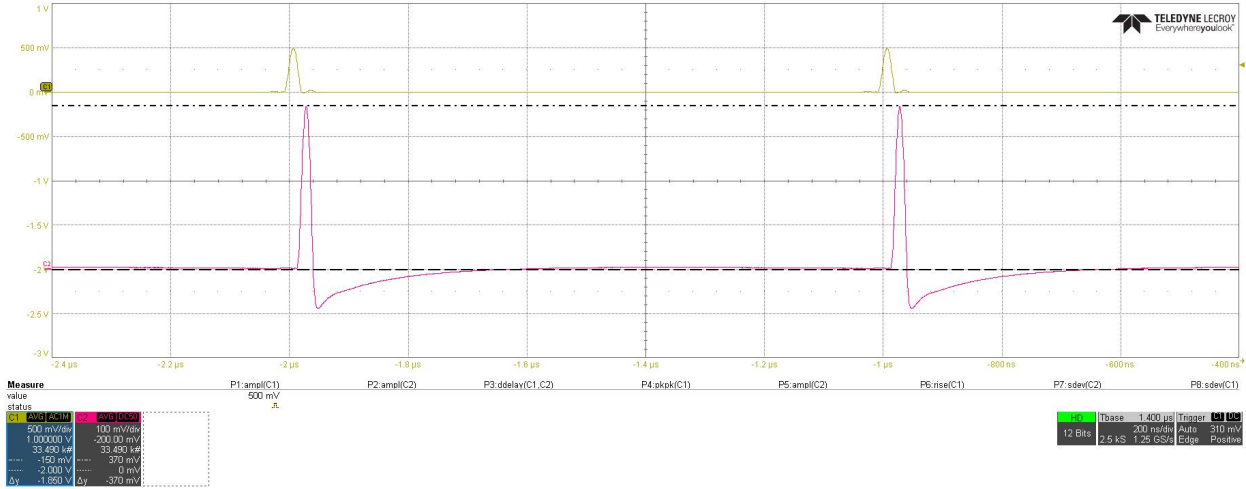

**Fig. 3.** Input (yellow) and output (magenta) waveforms used to characterize the electronics response with a calibrated current pulse source.

### 2.1.2. Avalanche signal

The PPC operates on the principle of avalanche ionization in gaseous media [13,16]. When a charged particle traverses the gas gap, it ionizes the $CO_2$ molecules along its trajectory, generating primary electron-ion pairs. Under a high electric field, the liberated electrons accelerate toward the anode and initiate secondary ionizations through collisions, resulting in electron-avalanche multiplication. This process is governed by the electron drift velocity ($v_e$) and the effective first Townsend coefficient (α), producing a short induced current pulse on the readout electrodes.

For this test the chamber was operated at a bias voltage of 1570 V, corresponding to an electric field of 39.25 kV/cm across the 0.4 mm gap (*g*). Under these conditions, the main swarm parameters were determined to be $v_e \approx$ 18.98 cm/µs, α ≈ 104.23 $cm^{-1}$, as computed using BOLSIG+ (version 03/2016) with the Biagi database [17, 18] (see Fig. 4).

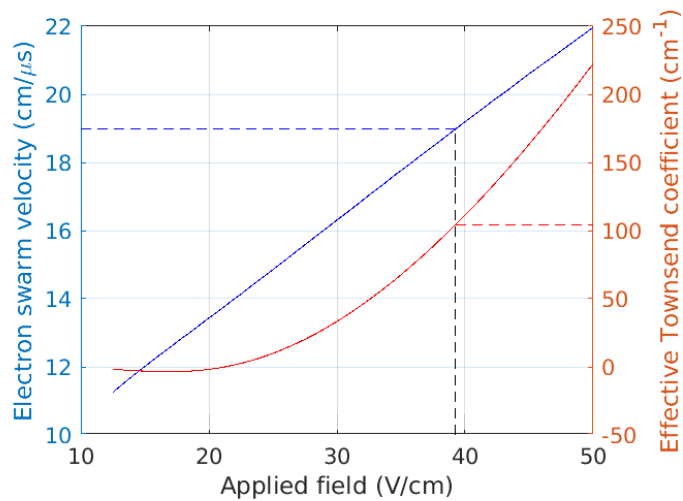

**Fig. 4.** Field-dependent electron drift velocity and Townsend coefficient as functions of the applied electric field in $CO_2$ gas (calculated from **[17, 18]**).

The cathode-to-anode gas gain $G_0$ can be estimated as:

$$G_0 = exp(\alpha g) = exp(104.23\ cm^{-1} \times 0.04\ cm) \approx 65$$

The electron transit time ($T_e$) across the gap is given by:

$$T_e = \frac{g}{v_e} = \frac{0.04\ cm}{18.98\ cm/\mu s} \approx 2.1\ ns$$

Neglecting diffusion, the idealized avalanche-induced current ($I_e$) is described by ([16], eq.4.31):

$$I_e = c(g - v_e t)exp(\alpha v_e t)$$

where $c$ is a proportionality constant. This function characterizes the temporal evolution of the induced current, rising sharply as the avalanche develops toward the anode.

The signal peaking time ($T_p$), defined as the time at which the induced current reaches its maximum, is given by:

$$T_p = T_e\left(1 - \frac{1}{\ln G_0}\right) \approx 1.6\ ns$$

Substituting $G_0 \approx 65$, the peaking time evaluates to approximately 1.6 ns. Note that in a real TOF measurement this delay will be identical in both chambers and will cancel out if the chambers are identical. Any deviations will be calibratable systematic errors.

## 2.2. Carbon Beam Experiment Condition and Data Acquisition

The experimental measurements were performed at the Yonsei Heavy-ion Therapy Center (Seoul, Republic of Korea) using a synchrotron-based carbon-ion radiotherapy system (CI-1000, Toshiba, Japan) [19]. A single PPC was precisely aligned with the beam axis in the fixed-beam room, as illustrated in Fig. 5. Detector signals were digitized and recorded using a Teledyne LeCroy HDO6104B digital oscilloscope (1 GHz bandwidth, 12-bit resolution). All acquired data were stored in binary format for offline processing.

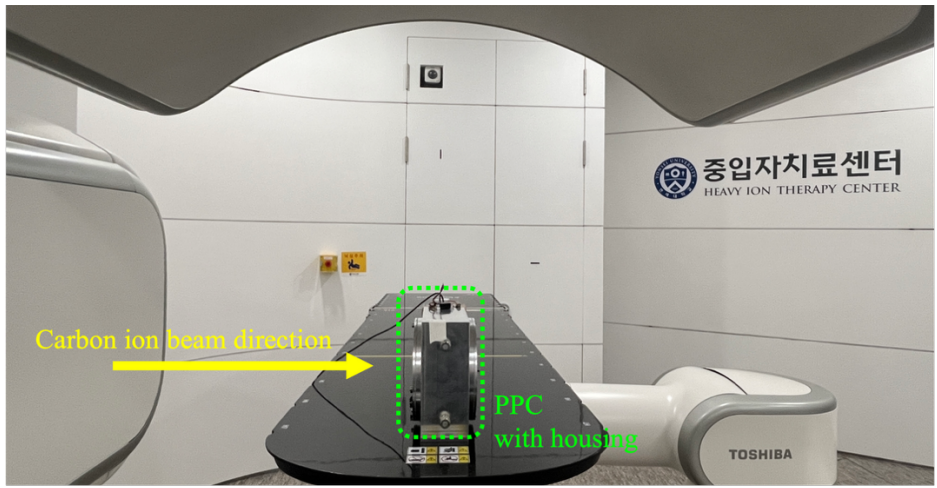

**Fig. 5.** Experimental PPC setup at the Yonsei CIRT center for evaluating detector timing performance under clinical carbon-ion beam conditions.

Four nominal beam energies were investigated: three clinically employed settings at 140 MeV/nucleon (MeV/u) and above, and an additional 102 MeV/u condition to probe the lower operational limit of the detector. For each energy, three particle-rate levels (particles/s) were selected, covering the clinically relevant range. The beam spot size at the PPC location varied with energy, ranging from approximately 2.1 mm to 4.4 mm in standard deviation (σ). These beam parameters-including energy, particle rate, and spot size-are summarized in Table 1.

For each beam condition (defined by energy and intensity), approximately 30 consecutive sweeps of 20 ms duration were acquired at a fixed sampling rate of 1.25 Gs/s (it was determined in preliminary tests that there is no advantage in using a higher sampling rate).

**Table 1.** Summary of beam condition used in the carbon-ion experiment.

| Nominal energy (MeV/u) | Orbital frequency (MHz) | Intensity | Average particle rate ($s^{-1}$) | Spot size at PPC (σ mm) |
|---|---|---|---|---|
| 102.56 | 2.05 | Low | 2.20E+07 | 4.40 |
| | | Medium | 1.24E+08 | |
| | | High | 2.34E+08 | |
| 140.44 | 2.34 | Low | 2.76E+07 | 3.65 |
| | | Medium | 1.56E+08 | |
| | | High | 2.94E+08 | |
| 250.25 | 2.91 | Low | 4.13E+07 | 2.55 |
| | | Medium | 2.34E+08 | |
| | | High | 4.40E+08 | |
| 430.00 | 3.45 | Low | 5.65E+07 | 2.12 |
| | | Medium | 3.20E+08 | |
| | | High | 6.03E+08 | |

The carbon ion beam extracted from the synchrotron exhibits a quasi-periodic time structure, characterized by discrete pulse intervals (spills). The orbital frequency, defined as the repetition rate of the circulating carbon ion bunches, is primarily determined by the beam energy and the geometry of the synchrotron. In this study, the orbital frequency was calculated based on a synchrotron perimeter of 63.4 m. As summarized in Table 1, the calculated orbital frequencies ranged from 2.05 MHz at 102.56 MeV/u to 3.45 MHz at 430.00 MeV/u.

Although the nominal synchrotron revolution frequency is on the order of several megahertz, the observed spill frequency is approximately 1 MHz—effectively one-third of the expected orbital frequency. This reduced fundamental frequency likely reflects the timing structure of the extraction process and suggests that particle extraction occurs once every three turns rather than on every revolution [20,21]. As shown in Fig. 6, the base width of each spill varies with beam energy, measured to be approximately 100 ns for 102 MeV/u and 40 ns for 430 MeV/u.

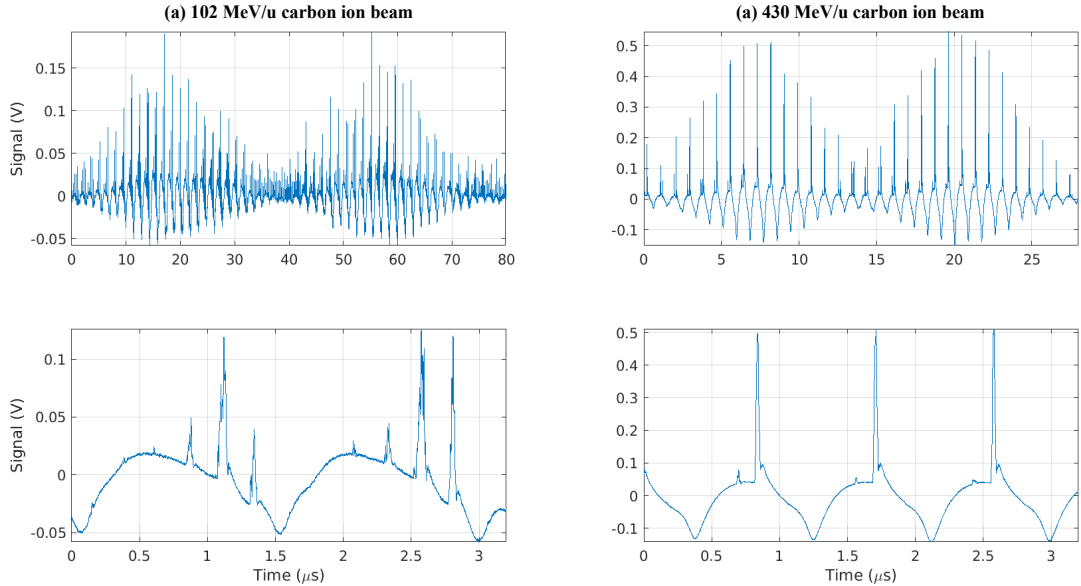

**Fig. 6.** Time-domain PPC current signals measured under different beam energies:
(a) 102 MeV/u, (b) 430 MeV/u
(lower row: magnified views showing individual spill structures).

## 2.3. Data Analysis

The raw waveform data acquired from the PPC was initially processed using a rectangular digital bandpass filter (10–200 MHz), chosen experimentally to optimize the timing performance. To estimate the intrinsic timing precision of the PPC, a synthetic time-of-flight (TOF) emulation method was employed using duplicated and time-shifted signals. Specifically, each acquired waveform was duplicated and temporally shifted by an integer number of samples, corresponding approximately to the fundamental period identified through frequency-domain analysis. This approach effectively generated two temporally offset signals that are independent of each other for time ranges much shorter than the offset time, mimicking the output of a dual-chamber TOF setup. This offset time (of the order of 1 µs) corresponds to a coarse fundamental period measurement, to be added to the fine time offset determined by the procedure described below.

Each 20 ms long acquired waveform (sweep) was further segmented into 10,000 sub-sweeps, each 2 µs in duration. For each sub-sweep, the following procedure was applied: (i) a 10-fold interpolation of both signals was performed to improve timing granularity, (ii) cross-correlation was computed over a predefined offset range, and (ii) the correlation peak was fitted with a parabola. The vertex of the fitted parabola was taken as the most likely time offset between the two signals, representing a single time measurement.

The overall analysis procedure is illustrated in Fig. 7. In panel (a), the second (orange) waveform is a temporally shifted replica of the first (blue) signal. Panel (b) shows the resulting cross-correlation function, and panel (c) presents a parabolic fit around the correlation peak. The vertex of the fitted parabola, marked by a green arrow, indicates the final time offset estimate used for timing precision evaluation.

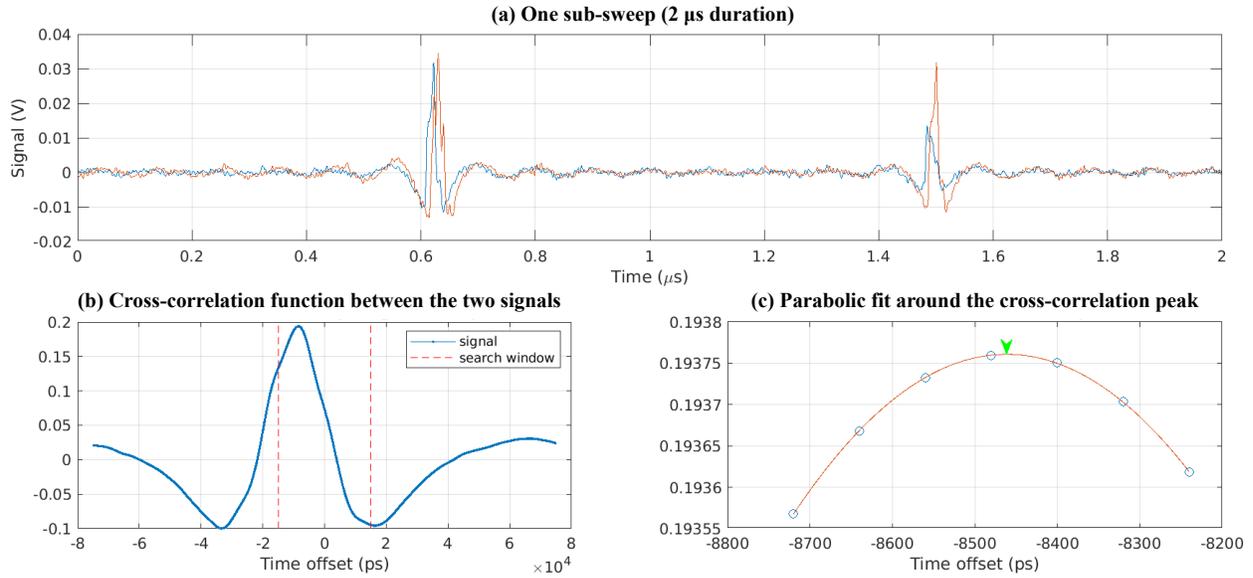

**Fig. 7.** Timing analysis procedure using synthetic TOF signals.
(a) One 2 μs long sub-sweep with duplicated and shifted waveforms, (b) Cross-correlation function between the two signals, and (c) Parabolic fit around the correlation peak; the vertex (green arrow) indicates the measured time offset.

To extract the overall timing precision for each beam energy, all time-offset measurements obtained at the three different beam intensities were aggregated and the distribution was fitted using a truncated logistic probability density function. As shown in Fig. 8, the fitted function reasonably described the observed distribution of time offsets. The timing precision was then estimated as one-quarter of the 95% confidence interval width of the location parameter (μ) obtained from the fit.

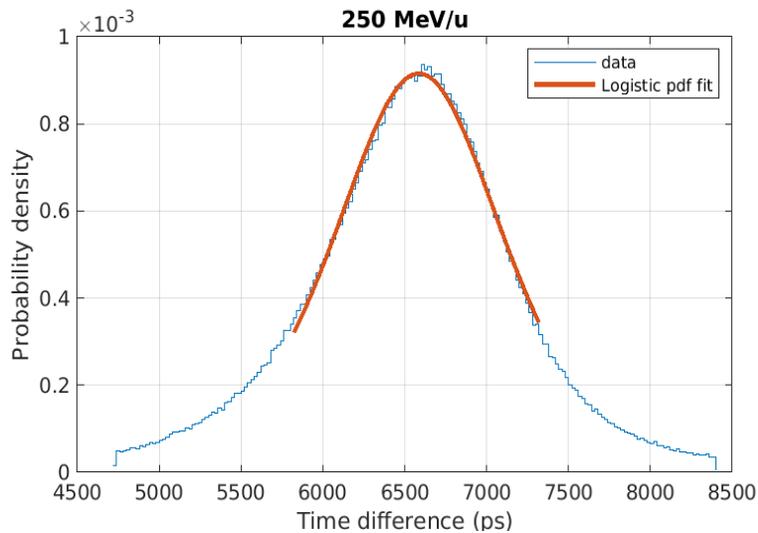

**Fig. 8.** Logistic PDF fit to the time difference distribution measured at 250 MeV/u.

# 3. Results

## 3.1. Stability of the PPC under Various Beam Conditions

The operational stability and current response of the PPC were evaluated under a range of carbon ion beam conditions, covering the clinically relevant settings. Through a voltage scan, a workable operating voltage of 1570 V was identified, at which the detector demonstrated sufficient sensitivity to achieve adequate timing precision, and no spark discharges were observed during operation.

As summarized in Table 2, the measured peak currents ranged from 0.73 µA to 7.58 µA, while the amplified signal root-mean-square (RMS) ranged from 1.42 mV to 14.99 mV, depending on the beam energy and intensity.

**Table 2.** Signal RMS and current of the PPC depends on the Carbon ion beam conditions

| Nominal energy (MeV/u) | Intensity | Signal RMS (mV) | Peak current (uA) |
|---|---|---|---|
| 102.56 | Low | 1.42 | 0.73 |
|  | Medium | 4.33 | 3.47 |
|  | High | 6.62 | 5.64 |
| 140.44 | Low | 1.91 | 0.81 |
|  | Medium | 7.28 | 3.85 |
|  | High | 11.4 | 6.21 |
| 250.25 | Low | 2.47 | 0.94 |
|  | Medium | 9.81 | 4.40 |
|  | High | 15.4 | 6.57 |
| 430.00 | Low | 2.90 | 1.07 |
|  | Medium | 12.9 | 4.95 |
|  | High | 15.0 | 7.58 |

## 3.2. Performance of the PPC with Synthetic TOF Emulation Analysis

To assess the intrinsic timing precision of the PPC detector, synthesized TOF signals were analyzed for four carbon-ion beam energies (102, 140, 250, and 430 MeV/u), each measured at three different beam intensities. The combined data per energy setting (as listed in Table 1) were used to calculate a single representative timing precision, summarized in Fig. 9.

The timing precision, defined as the standard deviation of the time offset measurements, was normalized to a one-second acquisition time to ensure comparability. This normalization abstracts the total sweep duration collected for each run and allows for extrapolation to other acquisition times via square root scaling.

Under this framework, the PPC demonstrated a few-picosecond precision across all beam energies, including the lowest-energy setting of 102 MeV/u. For reference, Fig. 9 also displays the expected TOF shift corresponding to a 1 mm shift in water-equivalent range, assuming a 30 cm time measurement baseline and using the clinical energy–range table employed at the Yonsei CIRT center. While the precision measured at 430 MeV/u marginally exceeds this reference threshold, all other energies fall well below it. These results confirm that the PPC detector is capable of resolving beam energy variations corresponding to sub-millimeter range deviations in water-equivalent depth.

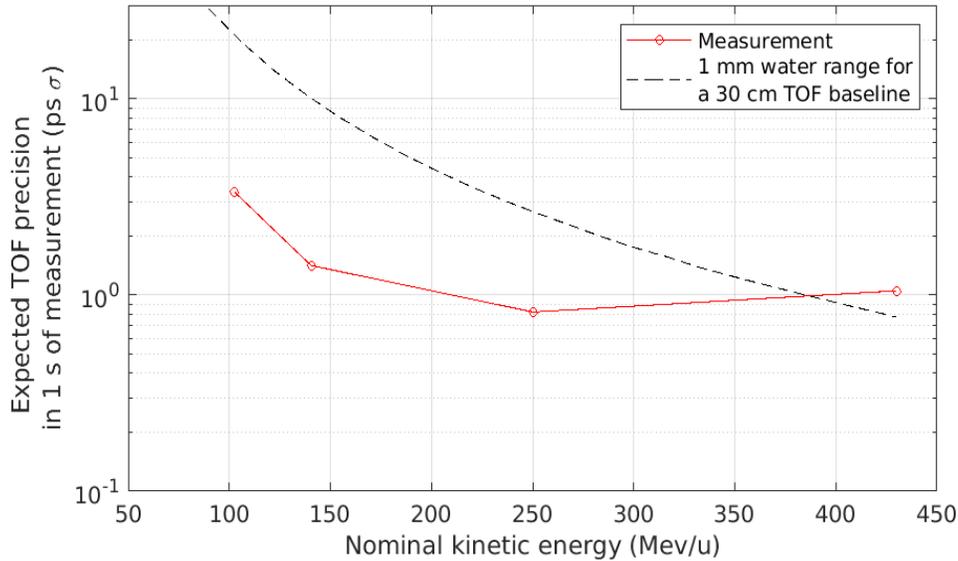

**Fig. 9.** PPC timing precision (1 s equivalent) vs. nominal beam energy.
(Dashed line: TOF precision for 1 mm range shift over a 30 cm baseline)

It is likely that for an actual TOF measurement between two chambers the precision will be somewhat better because in the present measurement there is no correlation between the details of the two waveforms, while for an actual TOF measurement there will be correlation to some extent as the same particle sequence will go through both chambers.

Fig. 10 presents the difference between the nominal kinetic energy and the beam energy inferred from the orbital frequency (see Table 1) measured using the method illustrated in Fig. 6. This includes the four carbon ion energies with all three beam intensities included per energy point. The dashed lines represent the energy deviations corresponding to a ±1 mm shift in water-equivalent range, as derived from the energy–range conversion tables used at the Yonsei CIRT facility. As the orbital perimeter is not exactly known it was determined by a global least-squares minimization of the energy discrepancy, yielding the value of 63.351 m.

Across all tested beam energies, the inferred kinetic energy deviation remained below 1 MeV. Notably, even at the highest energy of 430 MeV/u, the energy estimation remained within the ±1 mm range equivalence, confirming the PPC detector's capability for accurate orbital frequency-based energy measurement in therapeutic carbon ion beams.

The error bars, corresponding to Fig. 9, are invisible at the scale of the figure. The observed energy discrepancies correspond to hundreds of picosecond discrepancies in the fundamental period, so, in view of the precision determined above, it is not likely that these discrepancies have origin in the measurement but rather in the accelerator.

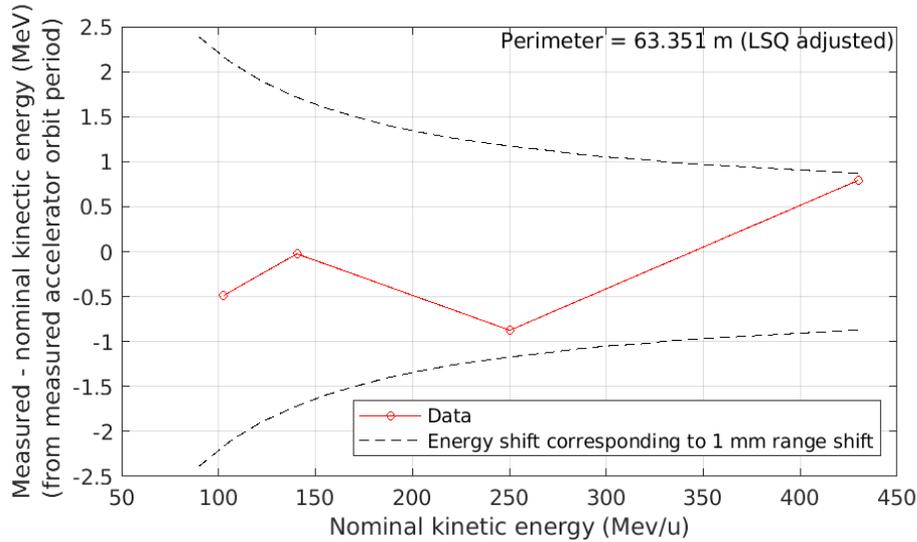

**Fig. 10.** Energy deviation from nominal values using orbital period measurement.
(Dashed lines: ±1 mm water range-equivalent energy shifts)

## 4. Discussion

In this study, we evaluated the performance of the PPC as a TOF detector for carbon ion beam monitoring. Our PPC demonstrated exceptional timing precision on the order of picoseconds, corresponding to a water-equivalent range resolution of approximately 1 mm—a critical requirement for clinical carbon ion therapy. As illustrated in Fig. 8, this high level of timing accuracy was consistently maintained across all four tested carbon-ion energies, indicating robust performance throughout the therapeutic energy range.

For precision and accuracy evaluation, we employed a synthetic TOF emulation technique using a single PPC detector. This approach successfully met the stringent clinical criteria for energy uncertainties (±1%) and range tolerance (±1 mm) [4]. Nonetheless, measurements at the highest tested energy of 430 MeV/u exhibited some limitations, attributable to the inherent limitations of using a single detector configuration. As depicted in Fig. 6 (a), the second (orange) waveform represents a time-shifted replica of the first (blue) signal within a 2 μs sub-sweep. The correlation analysis was performed between these artificially duplicated signals rather than between signals produced by the same particle sequence going through both chambers. This methodology is disadvantageous, as correlations are evaluated between fundamentally unrelated waveforms. We anticipate that an actual TOF measurement employing two independent PPC detectors will show enhanced timing precision.

Looking toward clinical integration, several enhancements to the current PPC design are imperative. The present PPC geometry, conceived for proof-of-concept validation, requires refinement for routine clinical deployment. Future developments should focus on implementing thinner electrodes to minimize beam energy degradation and expanding the active area to accommodate the full lateral scanning range of therapeutic beams. As a future study, implementation of a dual-PPC system will provide true delivery-point energy monitoring. This advancement could improve timing accuracy and enable real-time beam monitoring, ultimately contributing to enhanced treatment precision and patient safety.

## 5. Conclusion

The timing performance of the PPC detector was systematically evaluated under clinically relevant carbon-ion beam conditions, encompassing the full therapeutic range of energies and intensities. Notably, the PPC operated reliably throughout all tests without any occurrence of electrical sparking. Utilizing a synthesized TOF signal and a cross-correlation based algorithm, we demonstrated a timing precision approaching 1 picosecond for one-second-long acquisitions. This level of precision translates to a water-equivalent range uncertainty of approximately 1 mm, fully meeting the stringent accuracy requirements for clinical CIRT across all tested energies. Importantly, our findings suggest that true TOF measurements employing two independent PPC detectors are likely to yield even greater precision, owing to improved correlation between signals generated by the same particle sequence. The demonstrated compatibility of the PPC with $CO_2$—a non-polymerizing, radiation-hard gas—and the inherent inertness of the metallic electrodes further establish the chamber's suitability for high-rate, long-term operation in demanding clinical environments. Collectively, these results underscore the feasibility of PPCs as robust, radiation-hard detectors for real-time, time-based monitoring of CIRT beams. Future work will focus on implementing TOF measurements with dual, larger-area PPC chambers and extending the methodology to include proton beam applications. These advancements are expected to further enhance the precision and reliability of beam monitoring, supporting the ongoing evolution of particle therapy toward greater clinical efficacy and patient safety

## Acknowledgement

This work was supported by Korea Institute for Advancement of Technology (KIAT) grant funded by the government of the Republic of Korea (MOTIE, P0026103) and National Research Foundation of Korea (NRF) Grant funded by the Korean Government (RS-2024-00410034)## Acknowledgement

This work was supported by Korea Institute for Advancement of Technology (KIAT) grant funded by the government of the Republic of Korea (MOTIE, P0026103) and National Research Foundation of Korea (NRF) Grant funded by the Korean Government (RS-2024-00410034)

## References

[1] C.C. Ma, T. Lomax, Proton and Carbon Ion Therapy, CRC Press, Boca Raton, 2012.

[2] J.I. Kim, J.M. Park, H.G. Wu, Carbon ion therapy: a review of an advanced technology, Prog. Med. Phys. 31 (2020) 71–80.

[3] H. Tsujii, T. Kamada, T. Shirai, K. Noda, H. Tsuji, K. Karasawa, Carbon-Ion Radiotherapy, Springer, Tokyo, 2014.

[4] B. Arjomandy, P. Taylor, C. Ainsley, S. Safai, N. Sahoo, M. Pankuch, Y. Kase, AAPM task group 224: comprehensive proton therapy machine quality assurance, Med. Phys. 46 (2019) e678–e705.

[5] E.E. Klein, J. Hanley, J. Bayouth, F.F. Yin, W. Simon, S. Dresser, T. Holmes, Task Group 142 report: Quality assurance of medical accelerators, Med. Phys. 36 (2009) 4197–4212.

[6] S. Ishikura, Quality assurance of radiotherapy in cancer treatment: toward improvement of patient safety and quality of care, Jpn. J. Clin. Oncol. 38 (2008) 723–729.

[7] S. Jeong, M. Yoon, J. Son, Increased efficiency of range verification in routine QA for pencil-beam scanning proton therapy, J. Korean Phys. Soc. 73 (2018) 983–989.

[8]  M. Shimbo, E. Urakabe, Y. Futami, K. Yusa, H. Yamashita, N. Matsufuji, T. Kanai, Development of a multi-layer ion chamber for measurement of depth dose distributions of heavy-ion therapeutic beam for individual patients, Nippon Acta Radiol. 60 (2000) 274–279.

[9]  A. Vai, A. Mirandola, G. Magro, D. Maestri, E. Mastella, A. Mairani, M. Ciocca, Characterization of a MLIC detector for QA in scanned proton and carbon ion beams, Int. J. Part. Ther. 6 (2019) 50–59.

[10] Y. Yun, M.C. Han, C. Kim, C.S. Hong, D.W. Kim, H. Lee, J.S. Kim, Enhancing patient-specific quality assurance in carbon-ion radiation therapy: recalculating delivered dose distribution using log data, In Vivo 39 (2025) 1086–1093.

[11] Z. Chen, M.F. Moyers, Y. Deng, H.L. Chen, J. Li, Z.M. Shen, P. Yepes, Analysis of delivery and recalculation of dose using DICOM treatment records, Radiat. Med. Prot. 3 (2022) 123–130.

[12] E. Urakabe, M. Inoue, Y. Iwashita, T. Shirai, T. Sugimura, A. Noda, Y. Fujita, Performance of parallel plate ionization chamber for medical irradiation, in: Proc. Particle Accelerator Conf., IEEE, 1997, pp. 3819–3821.

[13] M. Abbrescia, V. Peskov, P. Fonte, Resistive Gaseous Detectors: Designs, Performance, and Perspectives, Wiley, 2018.

[14] N. Hamatani, M. Yagi, S. Shimizu, N. Ishino, M. Shimizu, Y. Kuwana, T. Kanai, Investigation of ionization chamber characteristics for ultrahigh-dose-rate scanned carbon-ion beams, In Vivo 38 (2024) 2220–2227.

[15] G. Rigoletti, R. Guida, B. Mandelli, Performance studies of RPC detectors operated with $C_2H_2F_4$ and $CO_2$ gas mixtures, Nucl. Instrum. Methods Phys. Res. A 1049 (2023) 168088.

[16] P. Fonte, Survey of physical modelling in Resistive Plate Chambers, J. Instrum. 8 (2013) P11001.

[17] BOLSIG+ solver ver. 03/2016, available at: https://www.lxcat.net, accessed on June 9, 2025.

[18] Biagi database, available at: https://www.lxcat.net, accessed on June 9, 2025.

[19] M.C. Han, S.H. Choi, C.S. Hong, Y.B. Kim, W.S. Koom, J.S. Kim, K.C. Keum, The first Korean carbon-ion radiation therapy facility: current status of the Heavy-ion Therapy Center at the Yonsei Cancer Center, Radiat. Oncol. J. 42 (2024) 295.

[20] S. Sato, T. Furukawa, K. Noda, Dynamic intensity control system with RF-knockout slow-extraction in the HIMAC synchrotron, Nucl. Instrum. Methods Phys. Res. A 574 (2007) 226–231. https://doi.org/10.1016/j.nima.2007.01.174

[21] T. Fujimoto, M. Kanazawa, T. Shirai, Y. Iwata, H. Uchiyama, K. Noda, Acceleration of heavy ions with a new RF system at HIMAC synchrotron, Nucl. Instrum. Methods Phys. Res. B 269 (2011) 2886–2890.